\documentclass[iop]{emulateapj}  

\usepackage{graphicx} 
\usepackage{apjfonts}

\shorttitle{OVERTURNING CONVECTION IN PENUMBRAL FILAMENTS}
\shortauthors{BELLOT RUBIO ET AL.}

\begin{document}

\title{Searching for overturning convection in penumbral filaments: 
slit spectroscopy at 0.2 arcsec resolution}

\author{L.R. Bellot Rubio$^1$}
\author{R. Schlichenmaier$^2$}
\author{K. Langhans$^3$}

\affil{$^1$Instituto de Astrof\'{\i}sica de Andaluc\'{\i}a (CSIC),
  Apdo.\ 3004, 18080 Granada, Spain}
\affil{$^2$Kiepenheuer-Institut f\"ur Sonnenphysik, Sch\"oneckstr. 6, 79104, 
Freiburg, Germany,}
\affil{$^3$Sickingenweg 10, 23568 L\"ubeck, Germany}

\slugcomment{To appear in ApJ. Accepted 2010 September 24}

\begin{abstract}

  Recent numerical simulations of sunspots suggest that overturning
  convection is responsible for the existence of penumbral filaments
  and the Evershed flow, but there is little observational evidence of
  this process. Here we carry out a spectroscopic search for
  small-scale convective motions in the penumbra of a sunspot located
  5$^\circ$ away from the disk center.  The position of the spot is
  very favorable for the detection of overturning downflows at the
  edges of penumbral filaments. Our analysis is based on measurements
  of the \ion{Fe}{1} 709.0~nm line taken with the Littrow spectrograph
  of the Swedish 1 m Solar Telescope under excellent seeing
  conditions.  We compute line bisectors at different intensity levels
  and derive Doppler velocities from them. The velocities are
  calibrated using a nearby telluric line, with systematic errors
  smaller than 150~m~s$^{-1}$.  Deep in the photosphere, as sampled by
  the bisectors at the 80\%-88\% intensity levels, we always observe
  blueshifts or zero velocities. The maximum blueshifts reach 1.2
  km~s$^{-1}$ and tend to be cospatial with bright penumbral
  filaments. In the line core we detect blueshifts for the most part,
  with small velocities not exceeding 300 m~s$^{-1}$.  Redshifts also
  occur, but at the level of 100-150 m~s$^{-1}$, and only
  occasionally. The fact that they are visible in high layers casts
  doubts on their convective origin.  Overall, we do not find
  indications of downflows that could be associated with overturning
  convection at our detection limit of 150~m~s$^{-1}$.  Either no
  downflows exist, or we have been unable to observe them because they
  occur beneath $\tau =1$ or the spatial resolution/height resolution
  of the measurements is still insufficient.

  \vspace*{1em}

\end{abstract}

\section{Introduction}

The origin of the Evershed flow is not completely understood
\citep{borrero09, scharmer09, schliche09, bellot10,
nordlundscharmer10}. Models based on moving flux tubes explain the
flow in terms of a pressure gradient that builds up along the tubes as
they rise from the sunspot magnetopause
\citep{schlicheetal98,schliche02}, whereas siphon flow models invoke
pressure gradients created by different field strengths at the
footpoints of elevated magnetic arches \citep[e.g.,][]{meyerschmidt68,
  degenhardt89,montesinosthomas97}. In both cases, mass conservation
is secured by downflows that occur in the mid and outer penumbra as
the field lines return to the solar surface.

Another possibility is that the Evershed flow is caused by convection
in the presence of the sunspot magnetic field, which induces an
anisotropy in the radial direction. The idea was put forward by
\citet{scharmeretal08} and seems to be supported by radiative
magnetohydrodynamic simulations \citep{heinemannetal07, rempeletal09a, 
rempeletal09b}.  According to the simulations, hot weakly
magnetized material ascends in the penumbra and becomes nearly
horizontal after being deflected outward by the inclined sunspot
field. This results in a penumbral filament ---an elongated
overturning flow pattern with an upward component at the center (the
Evershed flow) and lateral downflows at the edges. The process is
similar to granular convection in the quiet Sun, except for the
existence of a preferred horizontal direction.

In the simulations one observes narrow (0\farcs2--0\farcs4) lanes of
downflows on either side of the filaments. They appear where the
Evershed flow returns back to the solar surface.  The downflows ensure
mass conservation and reach velocities of up to 1.5 km~s$^{-1}$ at
optical depth $\tau=0.1$ \citep{heinemannetal07, rempeletal09a}.
Interestingly, they have a small horizontal component toward the
umbra, i.e., their direction is opposite to that of the Evershed flow.

A clear detection of downward motions at the edges of penumbral
filaments would support the existence of overturning convection in
sunspots. However, this is not an easy task because the Evershed flow is much
stronger and may hide them in an efficient way, especially at low
angular resolution. To minimize the problem, it is convenient to
investigate the sunspot regions perpendicular to the line of symmetry,
i.e., the line connecting the sunspot center with the disk center;
there, the Doppler shifts induced by radial flows are zero,
facilitating the detection of vertical motions. For an unambiguous
identification of downflows using Doppler measurements, the spot
should be as close to the disk center as possible, so that redshifts
can reliably be associated with downflows and blueshifts with upflows.

Despite these problems, there have been some reports of small-scale
convective motions in sunspot penumbrae. \citet{sanchezalmeidaetal07}
found a positive correlation between upward velocities and brightness
from high-resolution spectroscopic measurements taken at the Swedish
1m Solar Telescope (SST). The observed correlation is reminiscent of
normal convection, suggesting that a similar mechanism might be at
work in the penumbra.  \citet{rimmele08} studied penumbral flows near the
disk center using the Universal Birefringent Filter at the Dunn Solar
Telescope.  He detected upflows in the dark core and downflows on
either side of a penumbral filament, as expected from overturning
convection. However, other filaments in the same filtergrams did not
show downflows.

Using the Hinode sate\-llite, \citet{ichimotoetal07} discovered an
apparent twisting motion of brightness fluctuations in penumbral
filaments located at $\pm 90^\circ$ from the symmetry line. The
direction of the twist, as well as the associated Doppler signal
(albeit at lower angular resolution), were consistent with transverse
motions from the center of the filament to the edge facing the
observer. Similar conclusions have been obtained by
\citet{bhartietal10} from a larger sample of Hinode filtergrams. These
authors measured twist velocities of more than 2.1~km~s$^{-1}$ in the
portion of the filaments closer to the umbra and about 1~km~s$^{-1}$
at larger distances. \citet{zakharovetal08} reported the detection of
Doppler signals compatible with horizontal overturning motions in a
filament observed almost perpendicularly to the line of symmetry,
40$^\circ$ away from the disk center. Their measurements were taken
with the SOUP magnetograph at the SST. The line-of-sight velocity of
the horizontal motions was determined to be around 1~km~s$^{-1}$ from
a Milne-Eddington inversion of the data.  Unfortunately, the
\ion{Fe}{1} 630.25~nm line scans performed by Zakharov et al.\
required 123~s to be completed, meaning that any change in the flow
field, solar scenery, or even seeing conditions during the scan could
have affected the measurements.  Velocities of 1~km~s$^{-1}$ extending
over 0\farcs4 seem within easy reach and should have been detected
earlier.

By contrast, \citet{franzschliche09}  did not find indications
of overturning downflows in Doppler velocity maps computed from
spectropolarimetric observations taken with Hinode at a resolution of
0\farcs3. Also the theoretical work of \citet{borrerosolanki10} 
suggests that small-scale overturning motions are not needed to
explain the net circular polarization observed in the penumbra, since
the much stronger Evershed flow already accounts for it \citep[see][]{borreroetal07}.

In view of these results, the existence of overturning convection
cannot be considered as established.  Thus, it is important to
continue the search with measurements of the highest quality. Here we
present an analysis of sunspot observations made with the Littrow
spectrograph of the SST.  The spot was located very close to the disk
center, thus satisfying one of the requirements for good sensitivity
to downward motions mentioned above. These observations reach a
spatial resolution of about 0\farcs2 and were used
\citet{bellotetal05} to study the flow field of dark-cored penumbral
filaments.  To our knowledge, they are the highest spatial-resolution
spectroscopic measurements of sunspot penumbrae ever obtained. The
resolution is sufficient to distinguish the dark cores of penumbral
filaments, which are only barely detectable in the seeing-free
observations of the Hinode spectropolarimeter.

\section{Observations and data analysis}
\begin{figure}[t]
\begin{center}
\resizebox{.9\hsize}{!}{\includegraphics[bb=70 70 540 730]{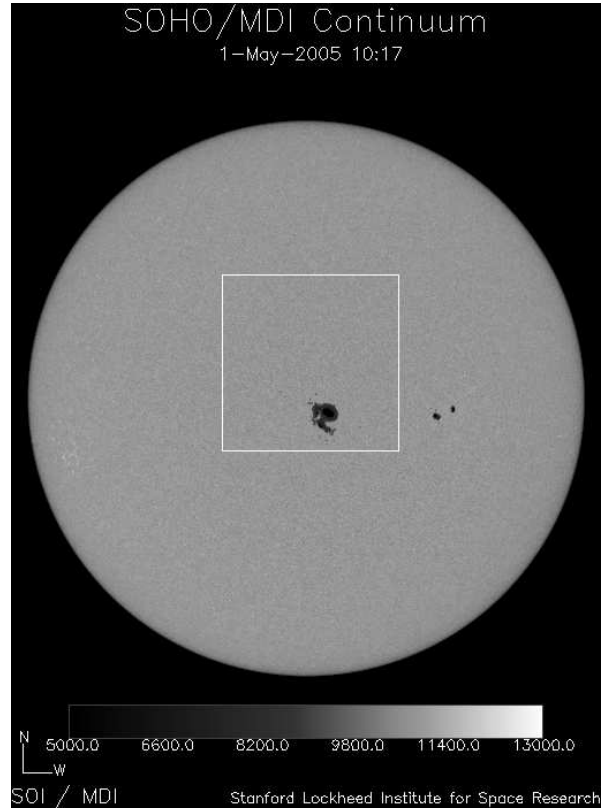}}
\end{center}
\caption{Full disk continuum image taken at 10:17 UT on 2005 May 1 by
  the Michelson Doppler Imager aboard SOHO. The spot closer to disk
  center is NOAA AR 10756, at coordinates S$6.9^\circ$ and
  W$4.3^\circ$.  Courtesy Stanford Lockheed Institute for Solar
  Physics.}
\label{mdi}
\vspace*{1em}
\end{figure}

\begin{figure*}[t]
\begin{center}
\resizebox{.73\hsize}{!}{\includegraphics[bb=100 80 1003 990,clip]{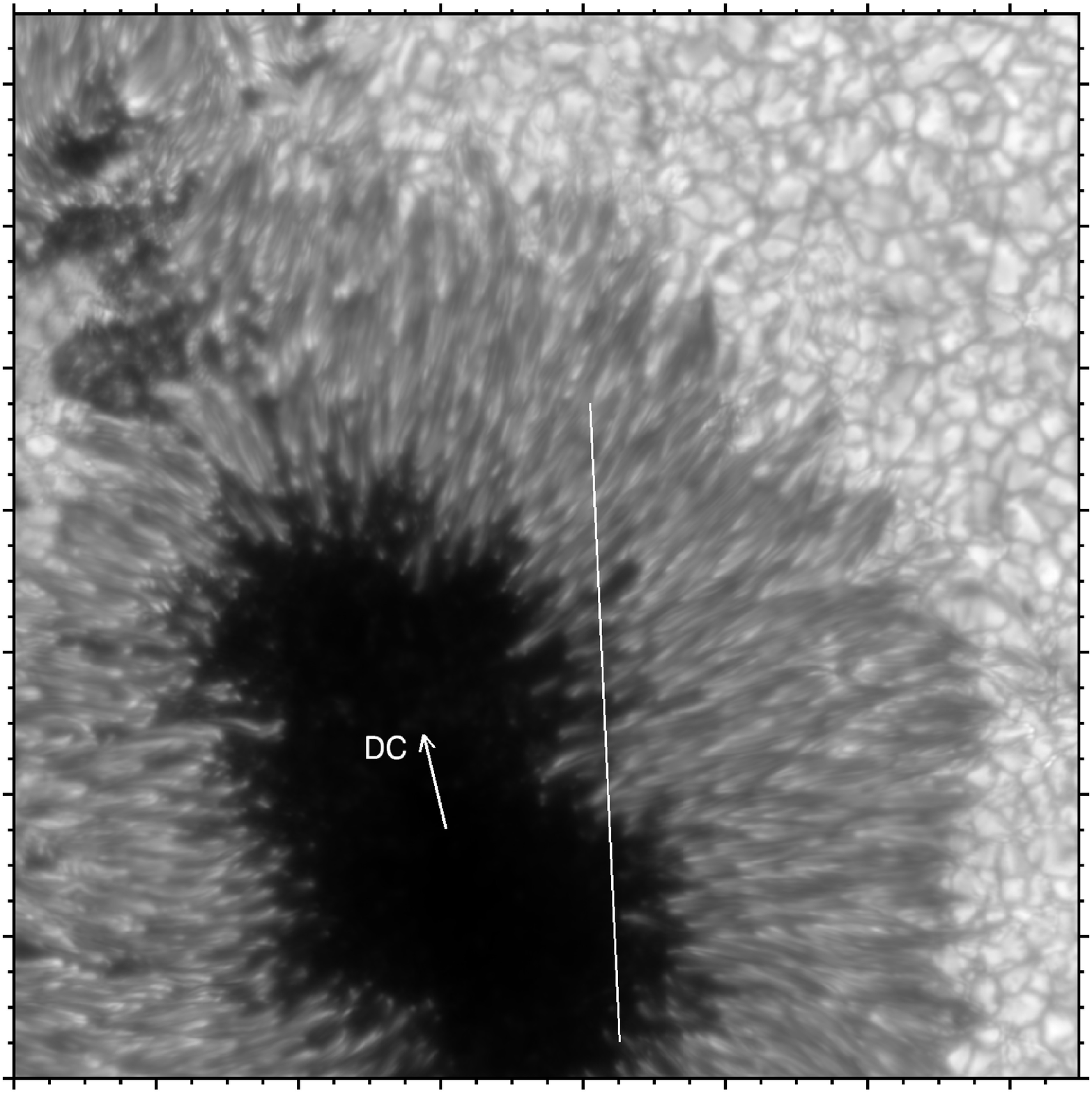}}
\resizebox{.255\hsize}{!}{\includegraphics[bb= 54 395 280 1000]{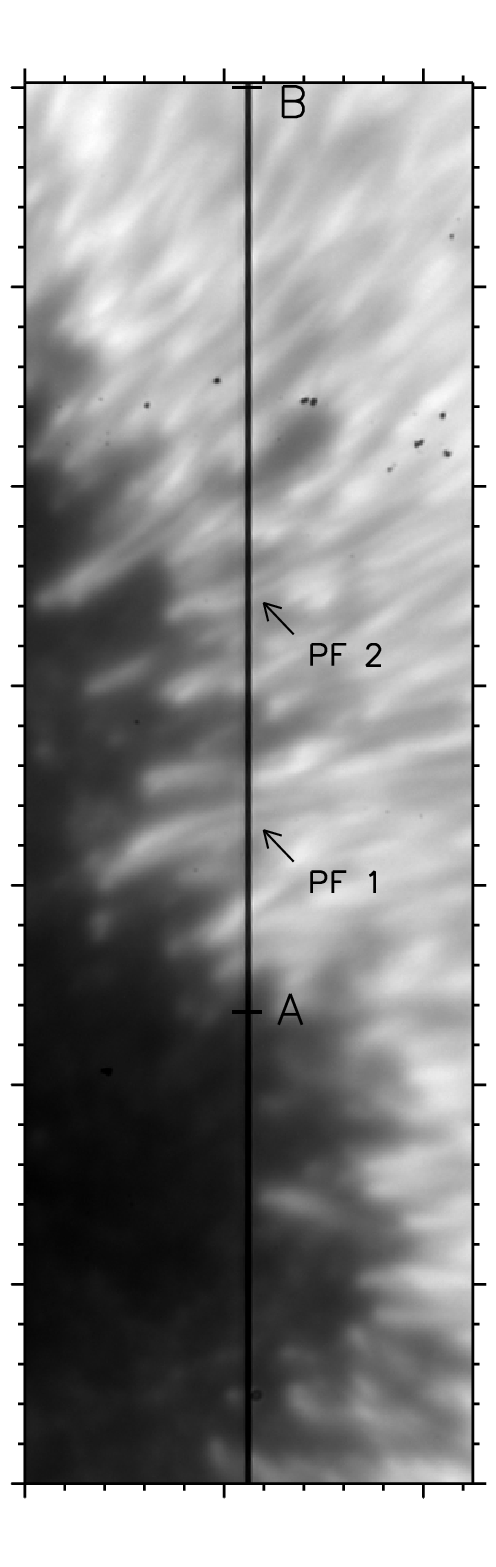}}
\end{center}
\caption{Left: G-band continuum image of AR 10756 obtained on 2005 May 1 at
  10:22 UT through a 1.1~nm wide filter centered at 436.4~nm. The image
  has been reconstructed from five individual filtergrams of  
  20 ms exposure time.  The arrow marks the direction to disk center (DC).
  The white line represents the spectrograph slit. It samples the
  umbra, the inner penumbra perpendicular to the symmetry line, and
  part of the center-side penumbra. Right: Slit-jaw image at 694~nm
  corresponding to the spectra discussed in this paper. The small
  dashes A and B indicate the portion of the slit shown in
  Figure~\ref{velocities}. Two dark-cored penumbral filaments (PF 1
  and PF 2) are clearly seen at the positions indicated by the
  arrows. }
\label{slit-jaw}
\vspace*{1em}
\end{figure*}

NOAA Active Region 10756 was observed with the Swedish 1 m Solar
Telescope \citep{scharmeretal03} on 2005 May~1. The best seeing
conditions occurred at around 10:22 UT, when the heliocentric angle of
the spot was 5$^\circ$. Figure~\ref{mdi} shows a full disk MDI
continuum image taken at 10:17 UT, while Figure~\ref{slit-jaw}
displays a context G-band continuum filtergram processed with the
Multi-Frame Blind Deconvolution technique \citep{vannoortetal05}. The
resolution of the G-band continuum image approaches 0\farcs1.

\subsection{Spectroscopic observations}

We used the TRI-Port Polarimetric Echelle-Littrow (TRIPPEL)
spectrograph and the adaptive optics system of the SST to obtain line
spectra at the highest angular resolution possible.  TRIPPEL\footnote{See 
http://dubshen.astro.su.se/wiki/index.php/TRIPPEL\_spectrograph} has a
grating of 79 grooves mm$^{-1}$, a blaze angle of 63.43$^\circ$, and a
theoretical resolving power of 240\,000.  \ion{Fe}{1} 557.6~nm,
\ion{Fe}{2} 614.9~nm, and \ion{Fe}{1} 709.0~nm were recorded
simultaneously during the observations, but in this paper we consider
only the 709.0~nm measurements because their short exposure times
provided the best angular resolution.

\ion{Fe}{1} 709.0~nm is a non-magnetic line suitable for Doppler shift
measurements. Its narrow shape makes it very sensitive to
line-of-sight velocities \citep{cabrerasolanaetal05}. We 
performed spectroscopic observations of the region around 709~nm with
a wavelength sampling of 1.046~pm, a pixel size of 0\farcs04, and an
exposure time of 200~ms. The spectra were recorded on a Kodak Megaplus
1.6i camera. The spectrograph slit had a width of 0\farcs11 (25
$\mu$m) and covered a length of 37\farcs8 (945 pixels).
Simultaneously, slit-jaw images were taken through a 1~nm wide filter
centered at 694~nm, also with exposure times of 200 ms.

The main advantage of slit spectrographs over tunable filters is that
they record the full line profile at once, thus preserving spectral
integrity. The disadvantage is that the solar image has to be stepped
across the spectrograph slit to create 2D maps. We performed
20\arcsec\/-wide scans of the spot, but only a few slit positions
attained a spatial resolution better than 0\farcs25. For this reason
we concentrate on the best slit position, not on the entire map.  As
can be seen in Figure 2, the slit crossed the center-side penumbra and
the region perpendicular to the line of symmetry near the umbra,
providing a wide range of positions where overturning convection could
occur.

The spectral images have been corrected for dark current, flatfield,
and optical distortions (smile and keystone). The latter produce
curvature of the spectral lines and a variation of the dispersion
along the slit, respectively. More details can be
found in \citet{langangenetal07}.

\subsection{Velocity measurements}

\begin{figure*}[t]
\begin{center}
\resizebox{1\hsize}{!}{\includegraphics[bb=63 380 838 643]{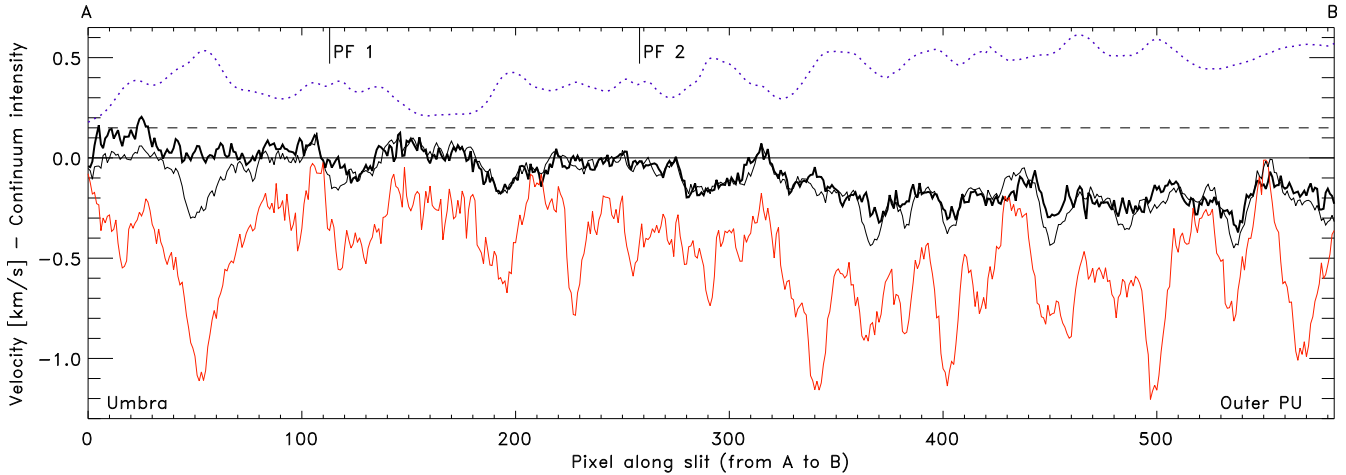}}
\end{center}
\caption{Absolute line-of-sight velocities derived from the line
  bisectors at different intensity levels: 0-8\% (thick black line),
  50-58\% (thin black line), and 80-88\% (red line). Positive
  velocities indicate redshifts. Since the spot was very near the disk
  center, they most likely correspond to downflows. The horizontal
  solid and dashed lines mark velocities of 0 and 150 m~s$^{-1}$,
  respectively. The dotted blue line displays the continuum intensity
  along the slit, in a relative scale. The vertical lines labeled PF1
  and PF2 in the upper abcissa mark the position of the two dark-cored
  penumbral filaments crossed by the slit (see Figure~\ref{slit-jaw}).}
\label{velocities}
\vspace*{.4em}
\end{figure*}

We derive absolute velocities from the observed spectra as follows.
First, we set up a relative wavelength scale using the dispersion
obtained from the pixel separation between two lines with
well-determined solar wavelengths in the corrected flatfield images.
To calibrate this scale in absolute terms, it is necessary to fix the
wavelength of one pixel in the spectrum. We use the telluric H$_2$O
line at 709.4050~nm for that purpose. The H$_2$O line is present in
the same spectral region and is not affected by motions in the solar
atmosphere, so it provides an excellent reference. Its central
wavelength has been obtained from the Fourier Transform Spectrometer
atlas of the quiet Sun by \citet{braultneckel87} with a precision
better than 50~m~s$^{-1}$. For each height along the slit, we
determine the position of the H$_2$O line core by means of a parabolic
fit. The derived values show pixel-to-pixel fluctuations of 110
m~s$^{-1}$ due to the weakness of the line, but we reduce them down to
40 m~s$^{-1}$ applying a 12-pixel boxcar average along the slit. 
The final result is one position in each spectrum with a very 
precise wavelength (that of the H$_2$O line), i.e., an absolute
wavelength scale.

The spectrally resolved profiles of \ion{Fe}{1} 709.0~nm are
used to derive Doppler shifts by means of line bisectors. We compute
bisectors for intensity levels from 0\% to 88\%, where 0\% represents
the line core and 100\% the continuum\footnote{We do not consider
intensity levels higher than 88\% to avoid the CN blend present in the
very far red wing of the Fe I line (at 709.069~nm).}, through linear
interpolation of the original wavelength samples. The bisector
positions are converted into Doppler shifts by subtracting the
laboratory wavelength of the
\ion{Fe}{1} line (709.03835~nm according to Nave et al.\ 1994), and
then transformed into line-of-sight velocities.

Different intensity levels sample different layers of the atmosphere.
Generally speaking, higher bisector intensities correspond to deeper
layers. In the penumbral atmosphere of \citet{bellotetal06},
the far line wing (as represented by intensities between 80\% and
88\%) is formed in the first 100 km above optical depth unity, i.e.,
the bisector Doppler shifts measured in this range reflect flow
velocities from the very deep photosphere.

The bisector velocities still need to be corrected for gravitational
redshift (636 m~s$^{-1}$) and relative motions between the Sun and the
observer. The former, computed as explained by
\citet{martinezpilletetal97}, amount to $+316$~m~s$^{-1}$, of which
$-261$~m~s$^{-1}$ correspond to the Earth's rotation,
$+436$~m~s$^{-1}$ to the Earth's orbital motion, and $+141$~m~s$^{-1}$
to the solar rotation at the position of the spot (negative velocities
are blueshifts).

Our {\em absolute} velocity calibration does not rely on the
granulation or the umbra, and therefore it is not affected by
uncertainties in the convective blueshift of the line or the presence
of umbral flows. The main source of systematic errors comes from the
laboratory wavelength of \ion{Fe}{1} 709.0~nm, which has an
uncertainty of about 100 m~s$^{-1}$ \citep{naveetal94}.  Another
source of systematic error is the central wavelength of the reference
H$_2$O line (50~m~s$^{-1}$). These errors add linearly, so the maximum
systematic error of our calibration is 150 m~s$^{-1}$.  The effect of
a systematic error is to shift all the velocities up or down as a
whole. In contrast, the pixel-to-pixel ''noise'' of the velocity
curves is due to random errors. We estimate the random error to be on
the order of 110 m~s$^{-1}$. This error comes from uncertainties in
the position of the reference H$_2$O line (40~m~s$^{-1}$) and
uncertainties in the determination of the bisector position (about 100
m~s$^{-1}$), added quadratically. All in all, our velocity
measurements should be accurate to within $\pm 110$ m~s$^{-1}$, with
systematic errors below 150 m~s$^{-1}$.

\section{Results}
Figure \ref{velocities} show the Doppler velocities observed along the
slit, excluding the dark umbra where it is not possible to compute
reliable positions for the weak H$_2$O line. The outer part of the
center-side penumbra is to the right. Displayed are line-core and
line-wing velocities corresponding to the bisector shifts averaged
between the 0\% and 8\% intensity levels (thick black line), between
50\% and 58\% (thin black line), and between 80\% and 88\% (red line).
As mentioned before, higher intensity levels progressively sample
deeper layers of the photosphere. For reference, the horizontal lines
indicate velocities of 0 and 150 m~s$^{-1}$ (positive values represent
redshifts). The continuum intensity is shown in the top part of the
figure to help identify bright and dark structures (dotted blue line).

The first thing to note from Figure \ref{velocities} is the tendency
of all the velocities to increase toward more blueshifted values as
the outer penumbra is approached, i.e., from left to right. This is
due to two reasons: the strong enhancement of the magnitude of the
Evershed flow with radial distance from the spot center (which
compensates the increasing inclination of the flow), and the more
favorable projection of the flow velocity to the line of sight toward
the upper end of the spectrograph slit. Indeed, the angle between the
penumbral filaments and the line of sight decreases from about
$90^\circ$ near the umbra to about 40$^\circ$ at the top of the slit;
since the Evershed flow occurs along the filaments, the projection
leads to stronger Doppler velocities. At an heliocentric angle
of only 5$^\circ$, however, the first effect is dominant.

\begin{figure*}[t]
\begin{center}
\resizebox{.24\hsize}{!}{\includegraphics[bb=67 370 387 949]{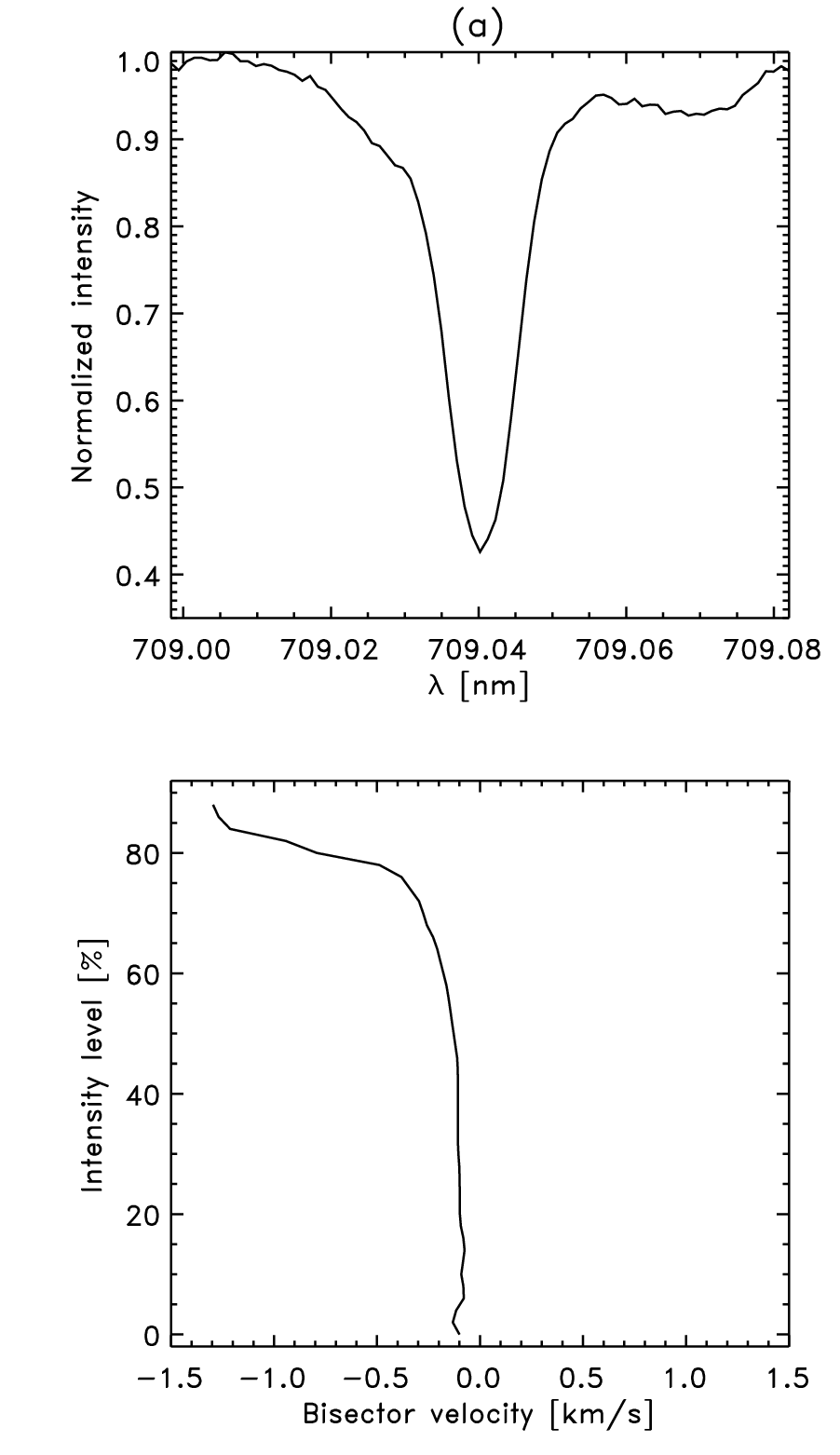}}
\resizebox{.24\hsize}{!}{\includegraphics[bb=67 370 387 949]{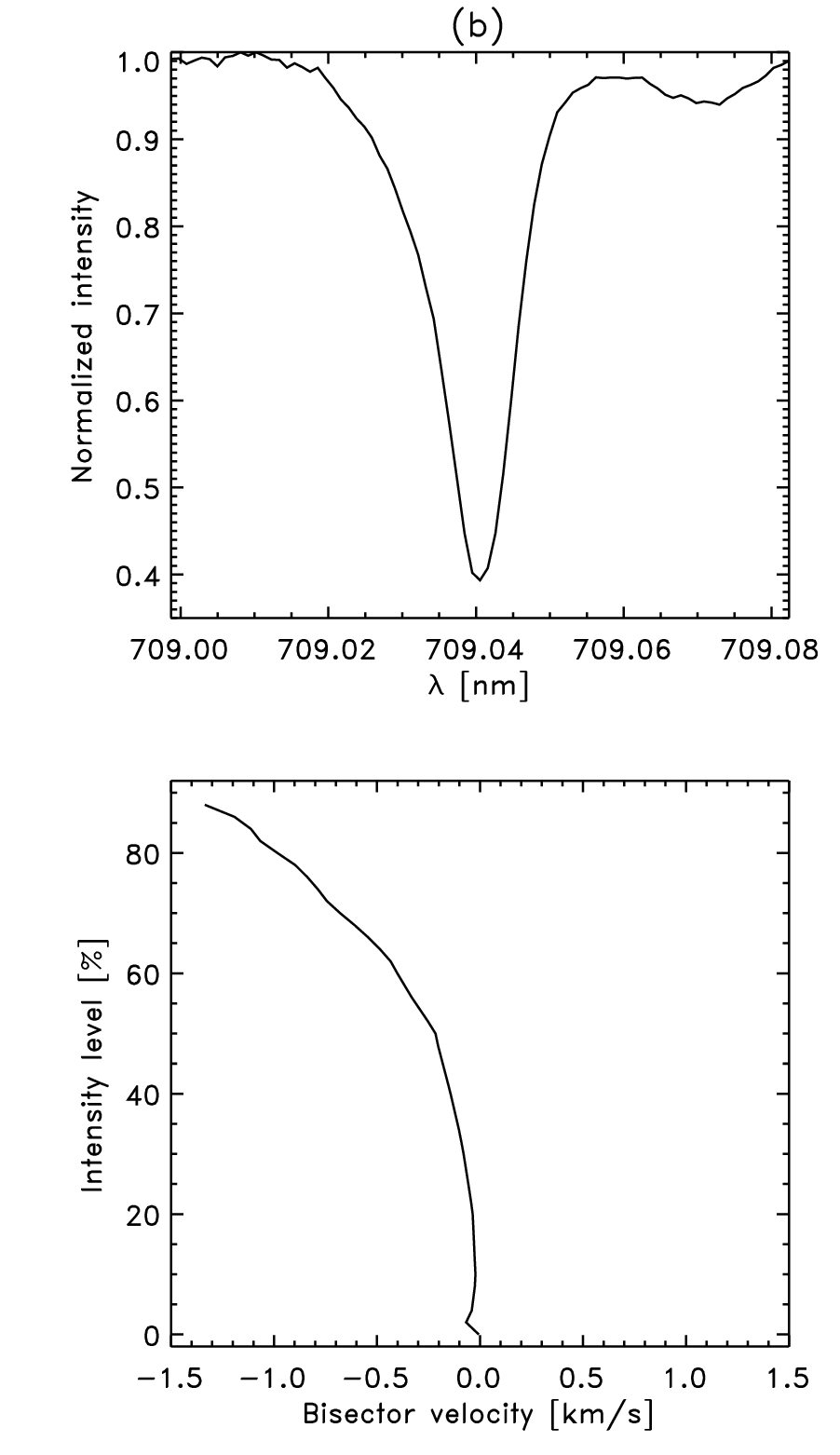}}
\resizebox{.24\hsize}{!}{\includegraphics[bb=67 370 387 949]{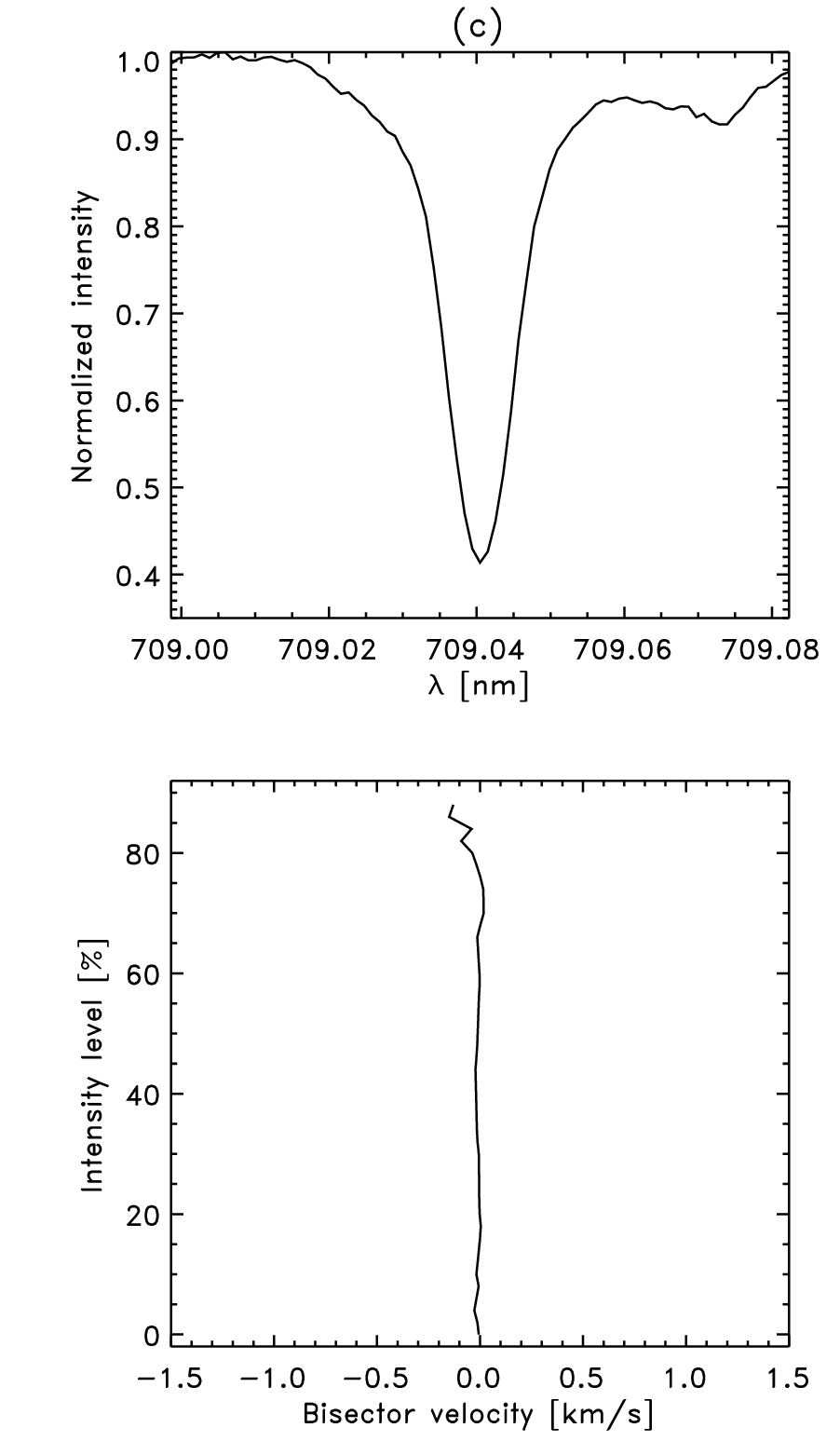}}
\resizebox{.24\hsize}{!}{\includegraphics[bb=67 370 387 949]{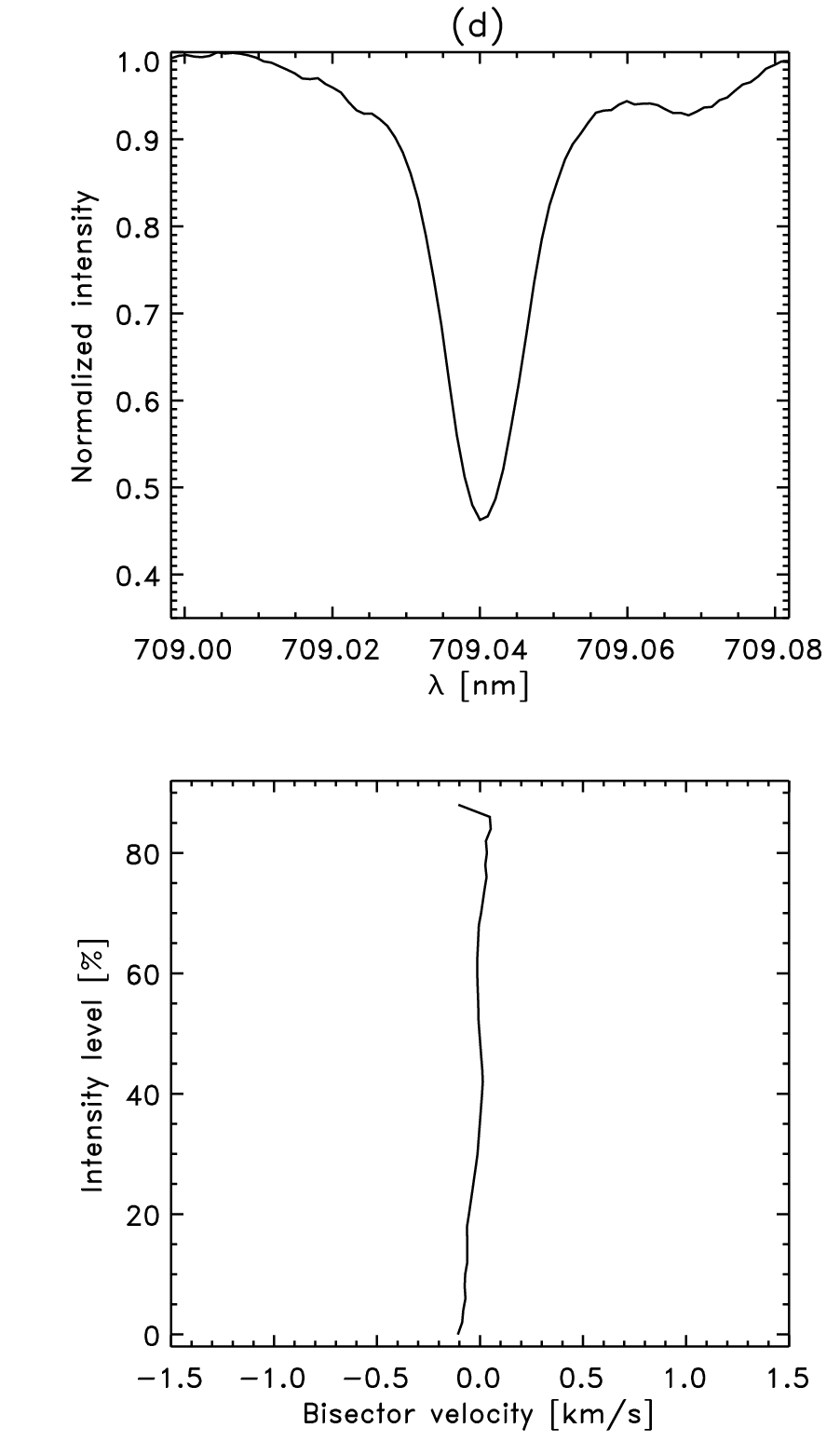}}
\end{center}
\caption{Examples of \ion{Fe}{1} 709.0~nm line profiles (top) and
  bisectors (bottom) at different positions along the slit. (a)
  Vertical bisectors with a strong tilt to the blue near the
  continuum. (b) Strongly tilted bisectors showing larger blueshifts
  toward the continuum. (c) Vertical bisectors with zero velocities.
  (d) Bisectors tilted to the red toward the continuum. These examples
  correspond to pixel positions 338, 54, 209, and 552 in Figure~3, respectively.} 
\label{bisectors}
\vspace*{1em}
\end{figure*}

We also note that the bisector velocities corresponding to intensity
levels of 0\%-8\% and 50\%-58\% are very similar. This means that the
lower half of the line show essentially vertical bisectors (i.e., no
line asymmetries). By contrast, the velocities are substantially
larger and to the blue in the far wings. Because of the larger
velocities at high intensity levels, the bisectors are tilted to the
blue near the continuum.  Figure~\ref{bisectors}(a) displays a typical
example: the strong tilt is caused by a ''satellite'' that starts to
be visible in the far blue wing of the line. Also depicted in the 
figure are bisectors showing a more gradual (but strong) shift
to the blue near the continuum, as well as cases of vertical bisectors
with zero velocities and bisectors showing a tilt to the red. The
latter, however, are not common in this part of the penumbra.

Deep in the atmosphere, as sampled by the bisectors near the
continuum, we always find blueshifts or zero velocities along the
slit. The maximum blueshifts reach 1.2 km~s$^{-1}$ and tend to be
cospatial with bright penumbral filaments, although we sometimes see a
small displacement of the maximum velocity away from the observer,
relative to the peak brightness position. In the dark regions outside
of the filaments we also detect blueshifts for the most part. However,
the velocities are strongly reduced. Sometimes they even drop to zero
(e.g., Figure~\ref{bisectors}c). Although these instances are not
common, this is the first time that zero velocities are observed in
dark areas of the center-side penumbra. Earlier measurements at lower
spatial resolution did not reveal them
\citep[e.g.,][]{hirzbergerkneer01, rouppe02, bellotetal06}. The blueshifts
depicted in Figure~\ref{velocities} may represent radial flows along
inclined flux tubes. In that case, one would expect very small or no
plasma motions in between the flow channels. However, this is not
observed. The relative absence of zero velocities in our data may
indicate that the tubes are still partly unresolved, or that there is
some amount of stray light contamination. Another possibility is that
the volume outside the flow channels is not completely at rest, as
suggested by Stokes inversions
\citep[e.g.,][]{bellotetal04,borreroetal05} and net circular
polarization measurements with Hinode \citep{ichimotoetal08}.

The line-core shifts (thick black curve in Figure~\ref{velocities})
are much smaller than their line-wing counterparts. They do not exceed
200 m~s$^{-1}$ except in the middle center-side penumbra. Most of the
positions along the slit show blueshifts. Redshifts do also occur, but
with velocities of less than 100-150 m~s$^{-1}$. These values are
smaller than the systematic errors of our absolute velocity
calibration. Apparently, the redshifts do not bear any relationship to
the penumbral filaments: sometimes they occur on one side of the
filament, sometimes on the other. The only exception is perhaps the
structure located at pixel position 20, which shows relatively
constant redshifts of about 100 m~s$^{-1}$.  This filament makes an
angle of about 90$^\circ$ to the symmetry line.

It is also important to note that the dark-cored penumbral
fi\-la\-ments crossed by the slit (PF 1 and PF 2 in
Figure~\ref{velocities}) do not exhibit blueshifts in the dark lanes
and redshifts in the two lateral brightenings, contrary to what would
be expected from a resolved overturning flow structure.

\section{Discussion}
The observations analyzed in this paper are very favorable for the
detection of downflows in the penumbra because they correspond to a
sunspot located only 5$^\circ$ away from disk center. The contribution
of horizontal flows to the observed Doppler signals is small, and
redshifts/blueshifts can safely be associated with vertical
downflows/upflows. Moreover, our observations provide full line
profiles at the highest angular resolution achieved in this kind of
measurements (better than 0\farcs25), including a telluric line that
we have employed to calibrate the velocity scale in absolute terms.

Despite the excellent quality of the data set, we do not detect
downflows that could be associated with overturning convection in deep
layers: near the continuum the measured velocities are always to the
blue. Higher in the atmosphere, as sampled by the line core, we
observe some regions that could harbor downflows, but with small
velocities not exceeding 150 m~s$^{-1}$. The fact that they are
visible only in high layers suggests that they are not related to
convective processes in the deep photosphere. They could be produced
by downflows associated with the inverse Evershed flow, but also by
penumbral oscillations or waves. In the absence of time series we
cannot decide between the different scenarios.

Overall, our data suggest that downflows due to overturning convection
are not larger than 150~m~s$^{-1}$ in the photospheric layers
accessible to the observations, while the simulations predict values
up to 1.5 km~s$^{-1}$. The lack of a clear detection of overturning
downflows may simply indicate that they do not exist. However, it is
also possible that they have gone undetected if (a) they occur beneath
$\tau = 1$; (b) they have velocities smaller than 150 m~s$^{-1}$; (c)
they are concentrated in very thin sheets not resolved by our
observations; or (d) they fill only a small fraction of the formation
region of the 709.0~nm line.

To examine possibility (c) we need spectroscopy at a resolution of
0\farcs1 or better. Even with infinite spatial resolution, (d) may
hamper the detection of the downflows if their vertical extent is much
smaller than the width of the contribution functions of typical
photospheric lines. In fact, the simulations show that the downflow
lanes surrounding the penumbral filaments are not completely
perpendicular to the solar surface, so that at some points they have a
thickness of only 50-100 km in the vertical direction \citep[e.g.,
Figure 9 of][]{rempeletal09a}. Such small structures may not be able
to leave clear signatures in the emergent intensity profiles. Line
synthesis calculations based on the simulation results are needed to
clarify this aspect. If \ion{Fe}{1} 709.0~nm does not provide
sufficient height resolution, then the search for overturning
downflows should be continued using lines with narrower contribution
functions, like \ion{C}{1} 538.03~nm. Another requirement to make
progress is to increase the realism of the simulations. Current
simulations reproduce many of the observed characteristics of the
penumbra, but they predict downflows that we do not detect in spite of
our excellent spatial, temporal, and spectral resolution. Thus, it is
important to confirm that their results are not affected by too large
values of the viscosity and magnetic diffusivity, by inadequate
boundary conditions, or by too shallow computational boxes. Hopefully,
the combination of these observational and numerical efforts will lead
to a better understanding of the penumbra and the Evershed flow.

\acknowledgements

Financial support by the Spanish Ministerio de Ciencia e Innovaci\'on
through project AYA2009-14105-C06-06 and by Junta de Andaluc\'{\i}a
through project P07-TEP-2687 is gratefully acknowledged.  The Swedish
1 m Solar Telescope is operated by the Institute for Solar Physics of
the Royal Swedish Academy of Sciences in the Spanish Observatorio del
Roque de los Muchachos of the Instituto de Astrof\'{\i}sica de Canarias.
This research has made use of NASA's Astrophysical Data System.


\begin{thebibliography}{}

\bibitem[Bharti et al.(2010)]{bhartietal10} Bharti, L., Solanki, 
S.~K., \& Hirzberger, J.\ 2010, ApJ, in press, arXiv:1009.2919 


\bibitem[Bellot Rubio(2010)]{bellot10} Bellot Rubio, L.R.\ 2010, in
  Astrophysics and Space Science Proceedings, Magnetic Coupling
  between the Interior and the Atmosphere of the Sun, ed. S. S. Hasan
  \& R. J. Rutten (Berlin: Springer), 193

\bibitem[Bellot Rubio et al.(2004)]{bellotetal04} Bellot Rubio, L.~R.,
  Balthasar, H., \& Collados, M.\ 2004, \aap, 427, 319

\bibitem[Bellot Rubio et al.(2005)]{bellotetal05}Bellot Rubio, L.R.,
  Langhans, K., Schlichenmaier, R.\ 2005, \aap, 443, L7

\bibitem[Bellot Rubio et al.(2006)]{bellotetal06} Bellot Rubio, L.~R., 
Schlichenmaier, R., \& Tritschler, A.\ 2006, \aap, 453, 1117 


\bibitem[Borrero(2009)]{borrero09} Borrero, J.~M.\ 2009, Science 
in China G, 52, 1670

\bibitem[Borrero et al.(2005)]{borreroetal05} Borrero, J.~M., Lagg,
  A., Solanki, S.~K., \& Collados, M.\ 2005, \aap, 436, 333

\bibitem[Borrero et al.(2007)]{borreroetal07} Borrero, J.~M., Bellot 
Rubio, L.~R., M\"uller, D.~A.~N.\ 2007, \apjl, 666, L133

\bibitem[Borrero \& Solanki(2010)]{borrerosolanki10} Borrero, J.~M.,
  \& Solanki, S.~K.\ 2010, \apj, 709, 349

\bibitem[Brault \& Neckel(1987)]{braultneckel87} Brault, J., \&
  Neckel, H. 1987, Spectral Atlas of Solar Absolute Disk-averaged and
  Disk-center Intensity from 3290 to 12510 \AA\/ (Hamburg: Hamb.
  Sternw.), ftp://ftp.hs.unihamburg.de/pub/outgoing/FTS-Atlas

\bibitem[Cabrera Solana et al.(2005)]{cabrerasolanaetal05} Cabrera
  Solana, D., Bellot Rubio, L.~R., \& del Toro Iniesta, J.~C.\ 2005,
  \aap, 439, 687


\bibitem[Degenhardt(1989)]{degenhardt89} Degenhardt, D.\ 1989,
  \aap, 222, 297


\bibitem[Franz \& Schlichenmaier(2009)]{franzschliche09} Franz,
  M., \& Schlichenmaier, R.\ 2009, \aap, 508, 1453


\bibitem[Heinemann et al.(2007)]{heinemannetal07} Heinemann, T., 
Nordlund, {\AA}., Scharmer, G.~B., \& Spruit, H.~C.\ 2007, \apj, 669, 1390 

\bibitem[Hirzberger \& Kneer(2001)]{hirzbergerkneer01} Hirzberger,
  J., \& Kneer, F.\ 2001, \aap, 378, 1078


\bibitem[Ichimoto et al.(2007)]{ichimotoetal07} Ichimoto, K., et al.\ 
2007, Science, 318, 1597 

\bibitem[Ichimoto et al.(2008)]{ichimotoetal08} Ichimoto, K., et al.\
  2008, \aap, 481, L9

\bibitem[Langangen et al.(2007)]{langangenetal07} Langangen, {\O}.,
  Carlsson, M., Rouppe van der Voort, L., \& Stein, R.~F.\ 2007, \apj,
  655, 615

\bibitem[Nave et al.(1994)]{naveetal94} Nave, G., Johansson, S.,
  Learner, R.~C.~M., Thorne, A.~P., \& Brault, J.~W.\ 1994, \apjs, 94,
  221


\bibitem[Nordlund \& Scharmer(2010)]{nordlundscharmer10} Nordlund,
  {\AA}., \& Scharmer, G.~B.\ 2010, in Astrophysics and Space
  Science Proceedings, Magnetic Coupling between the Interior and the
  Atmosphere of the Sun, ed. S. S. Hasan \& R. J. Rutten (Berlin:
  Springer), 243


\bibitem[Mart\'{\i}nez Pillet et al.(1997)]{martinezpilletetal97}
  Mart\'{\i}nez Pillet, V., Lites, B.~W., \& Skumanich, A.\ 1997,
  \apj, 474, 810

\bibitem[Meyer \& Schmidt(1968)]{meyerschmidt68} Meyer, F., \&
  Schmidt, H.~U.\ 1968, Zeitschrift Angewandte Mathematik und
  Mechanik, 48, 218

\bibitem[Montesinos \& Thomas(1997)]{montesinosthomas97} Montesinos,
  B., \& Thomas, J.~H.\ 1997, \nat, 390, 485


\bibitem[Rempel et al.(2009a)]{rempeletal09a} Rempel, M., 
Sch{\"u}ssler, M., \& Kn{\"o}lker, M.\ 2009a, \apj, 691, 640 

\bibitem[Rempel et al.(2009b)]{rempeletal09b} Rempel, M., Sch\"ussler, 
M., Cameron, R.~H., \& Kn\"olker, M.\ 2009b, Science, 325, 171 

\bibitem[Rimmele(2008)]{rimmele08} Rimmele, T.\ 2008, \apj, 672, 684

\bibitem[Rouppe van der Voort(2002)]{rouppe02} Rouppe van der Voort,
  L.~H.~M.\ 2002, \aap, 389, 1020


\bibitem[S{\'a}nchez Almeida et al.(2007)]{sanchezalmeidaetal07}
  S{\'a}nchez Almeida, J., M{\'a}rquez, I., Bonet, J.~A., \&
  Dom{\'{\i}}nguez Cerde{\~n}a, I.\ 2007, \apj, 658, 1357

\bibitem[Scharmer(2009)]{scharmer09} Scharmer, G.~B.\ 2009, Space
  Science Reviews, 144, 229

\bibitem[Scharmer et al.(2003)]{scharmeretal03} Scharmer, G.~B., 
Bjelksjo, K., Korhonen, T.~K., Lindberg, B., 
\& Petterson, B.\ 2003, \procspie, 4853, 341

\bibitem[Scharmer et al.(2008)]{scharmeretal08} Scharmer, G.~B., 
Nordlund, {\AA}., \& Heinemann, T.\ 2008, \apjl, 677, L149 


\bibitem[Schlichenmaier(2002)]{schliche02} Schlichenmaier, R.\ 
2002, Astronomische Nachrichten, 323, 303

\bibitem[Schlichenmaier(2009)]{schliche09} Schlichenmaier, R.\ 
2009, Space Science Reviews, 144, 213 


\bibitem[Schlichenmaier et al.(1998)]{schlicheetal98} 
Schlichenmaier, R., Jahn, K., \& Schmidt, H.~U.\ 1998, \aap, 337, 897

\bibitem[van Noort et al.(2005)]{vannoortetal05} van Noort, M., Rouppe
  van der Voort, L., L\"ofdahl, M.~G.\ 2005, \solphys, 228, 191

\bibitem[Zakharov et al.(2008)]{zakharovetal08} Zakharov, V.,
  Hirzberger, J., Riethm{\"u}ller, T.~L., Solanki, S.~K., \& Kobel,
  P.\ 2008, \aap, 488, L17

\end{thebibliography}
\end{document}